\documentstyle[12pt]{report}
\begin{document}
\begin{centering}
%\huge\bf Particle-number-conserving solution \\ of the generalized pairing problem
\huge\bf Exact solution for generalized pairing
\vskip .6truecm
{\normalsize\bf Feng Pan$^{\dagger}$ and J. P. Draayer}
\vskip .2cm
\noindent{\small\it Department of Physics {\normalsize\&} Astronomy,
Louisiana State University,}\\
{\small\it Baton Rouge, LA 70803-4001}\\

\vskip 1truecm
{\small\bf Abstract}\\
\end{centering}
\vskip .5cm
\normalsize An infinite dimensional algebra, which is useful for deriving exact solutions
of the generalized pairing problem, is introduced. A formalism for diagonalizing the
corresponding Hamiltonian is also proposed. The theory is illustrated with some numerical
examples.
\vskip 6cm
\noindent {-----------------------------------------}\\
\noindent $^{\dagger}$On leave from the Department of Physics,
Liaoning Normal University, Dalian 116029, P. R.~China

\newpage

   Pairing has long been considered an important interaction in physics. The concept can
be traced back to the seniority scheme introduced by Racah in atomic physics.$^{1}$ Its
physical significance was first realized in the study of superconductivity.$^{2}$
Following the suggestions of Bohr, Mottelson, and Pines,$^{3}$ the first detailed
application of pairing in nuclei was made by Belyaev.$^{4}$ The concept has since been
applied to other phenomena:  high T$_c$ superconductivity,$^{5,6}$ applications using
the Hubbard model,$^{7}$ pairing phenomena in liquids,$^{8}$ and metal clusters.$^{9}$
\vskip .3cm
BCS methods have yielded major successes in studies of superconductivity. When applied
to nuclei, however, some negatives come with the positives. First of all, not only is
the number of nucleons in a nucleus typically small, the number of valence
particles ($n\sim 10$) which dominates the behaviour of low-lying states is too few to
support underlying assumptions of the BCS approximation, specifically, $\delta n/n$ is not
negligible. As a consequence, particle-number-nonconservation effects enter and can lead
to serious difficulties, such as spurious states, nonorthogonal solutions, etc. Secondly,
an essential feature of pairing correlations are even-odd differences, which are driven
mainly by Pauli blocking. It is difficult to treat these differences in the BCS formalism
because different quasi-particle bases must be introduced for different blocked levels.
For these reasons, a particle-number-conserving method for treating the pairing problem
has been suggested for well-deformed nuclei.$^{10}$ The method uses a configuration
energy truncation scheme, and takes the strength of the pairing interaction to be the
same for all orbitals. In this limit the pairing Hamiltonian can be diagonalized in 
truncated configuration spaces. Because the theory applies to well-deformed nuclei,
each orbital can only be occupied by a single pair of particles. 
\vskip .3cm
   The generalized pairing Hamiltonian for spherical nuclei can be written as
\vskip .3cm
$$\hat{H}=
\sum_{jm}
\epsilon_{j}a^{\dagger}_{jm}a_{jm}-
\vert G\vert
\sum_{jj^{\prime}}c_{jj^{\prime}}S^{+}(j)S^{-}(j),\eqno(1)$$
\vskip .3cm
\noindent where the $\epsilon_{j}$ are single-particle energies and $S^{\pm}(j)$ and
$S^{0}(j)$ are the pairing operators for a single-$j$ shell defined by
\vskip .3cm
$$S^{+}(j)=\sum_{m>0}(-)^{j-m}a^{\dagger}_{jm}a^{\dagger}_{j-m},$$

$$S^{-}(j)=\sum_{m>0}(-)^{j-m}a_{j-m}a_{jm},$$

$$S^{0}(j)={1\over{2}}\sum_{m>0}(a^{\dagger}_{jm} a_{jm}+a^{\dagger}_{j-m} a_{j-m}-1)~=
~{1\over{2}}(\hat{N}_{j}-\Omega_{j}),\eqno(2)$$
\vskip .3cm
\noindent  where $\Omega_{j}\equiv j+1/2$ is the maximum number of pairs in the $j$-th
shell, $\hat{N}_{j}$ is the particle number operator for the $j$-th shell, and the 
$c_{jj^{\prime}}$ measure the pairing strength between different $j$-shells. In general,
for $N$  pairs, Hamiltonian (1) can be diagonalized in bases states that are products
of the single-$j$ shell pairing wave functions:
\vskip .3cm
$$\vert N>=\sum_{k_{i}}B_{k_{1}k_{2}\cdots
k_{p}}S^{+~k_{1}}_{j_{1}}S^{+~k_{2}}_{j_{2}}S^{+~k_{3}}_{j_{3}}\cdots 
S^{+~k_{p}}_{j_{p}}\vert 0>,\eqno(3)$$
\vskip .3cm
\noindent where the summation is restricted by 
\vskip .3cm
$$\sum_{i=1}^{p}k_{i}=N, \eqno(4)$$
\vskip .3cm
\noindent the $B_{k_{1}k_{2}\cdots k_{p}}$ are expansion coefficient that need to be
determined, and $\vert 0>$ is the pairing vacuum state which satisfies the condition
\vskip .3cm
$$S^{-}_{j}\vert 0>~=~0~~~{\rm for~ all}~~j.\eqno(5)$$
\vskip .3cm
\noindent The dimensionality of the Hamiltonian matrix in this basis increases very rapidly
with increasing $N$ and the number of shells $p$. It is less than or equal to the dimension of the
irreducible representation (irrep) $[N\dot{0}]$ of the unitary group $U(p)$ due to the Pauli principle, 
\vskip .3cm
$$\dim~\leq~{(p+N-1)!\over{N!(p-1)!}}.\eqno(6)$$
\vskip .3cm
\noindent The equal sign holds in (6) when all the single-j shell pairing wave functions in the
summation of (3) are Pauli allowed.
Indeed, the problem quickly becomes intractable because there are no analytical
expressions or recursion relations for determining the $B_{k_{1}k_{2}\cdots k_{p}}$
coefficients.
\vskip .3cm
   Following the quasi-spin approximation,$^{11}$ consider a simpler Hamiltonian 
\vskip .3cm
$$\hat{H}=-\vert G\vert S^{+}_{0}S^{-}_{0},\eqno(7)$$
\vskip .3cm
\noindent where
\vskip .3cm
$$S^{+}_{0}=\sum_{j}c^{*}_{j}S^{+}(j),~S^{-}_{0}=\sum_{j}c_{j}S^{-}(j)\eqno(8)$$
\vskip .3cm
\noindent with the coefficient $c_{i},~c^{*}_{j}$ satisfying the condition
\vskip .3cm
$$\sum_{i}\vert c_{i}\vert^{2}=1.\eqno(9)$$
\vskip .3cm
\noindent This defines generalized pairing as proposed by Talmi.$^{12}$ Clearly, in the
notation of (1), $c_{jj^{\prime}}=c^{*}_{j}c_{j^{\prime}}$ with $\vert c_{j}\vert^{2}$
giving the percentage of single-$j$ shell pairing in the Hamiltonian. In what follows, the
$c_{j}$ are taken to be real.
\vskip .3cm
To diagonalize Hamiltonian (7), consider an algebra generated by
\vskip .3cm
$$S^{0}_{m}=\sum_{j} c_{j}^{2m}S^{0}(j),$$

$$S^{\pm}_{m}=\sum_{j}c_{j}^{2m+1}S^{\pm}(j).\eqno(10)$$
\vskip .3cm
\noindent  It is easy to show that these generators satisfy the following
commutation relations:
\vskip .3cm
$$[S^{+}_{m},~S^{-}_{n}]~=~2S^{0}_{m+n+1},$$

$$[S^{0}_{m},~S^{\pm}_{n}]~=~\pm S^{\pm}_{m+n}.\eqno(11)$$
\vskip .3cm
\noindent Therefore, the $\{ S^{\mu}_{m},~\mu=0,+,-;~m=0,\pm 1, \pm 2,\cdots\}$ form an
infinite-dimensional algebra, one that differs only slightly from a general Lie
algebra of the affine type.  
\vskip .3cm
  The unique lowest-weight state of this algebra is simply the product of the single-$j$
shell pairing vacua with arbitrary seniority quantum numbers. Therefore, it suffices to 
consider the total seniority zero case. The lowest-weight state satisfies
\vskip .3cm
$$S^{-}_{m}\vert 0>=0;~m=0,~\pm 1,~\pm 2,~\cdots,\eqno(12)$$
\vskip .3cm
\noindent and
\vskip .3cm
$$S^{0}_{m}\vert 0>=-{1\over{2}}\sum_{j}\vert c_{j}\vert^{2m}\Omega_{j}\vert 0>=
\Lambda_{m}\vert 0>.\eqno(13).$$
\vskip .3cm
\noindent Furthermore, it can be proven that the eigenvectors of $\hat{H}$ for any $N$ and non-zero
energy eigenvalue  can be expanded as
\vskip .3cm
$$\vert N>=\sum_{n_{i}}x_{1}^{n_{1}}x_{2}^{n_{2}}\cdots
x^{n_{N-1}}_{N-1}S^{+}_{0}S^{+}_{n_{1}}S^{+}_{n_{2}}\cdots S^{+}_{n_{N-1}}\vert 0>,\eqno(14)$$
\vskip .3cm
\noindent where 
\vskip .3cm
$$n_{i}~=\left\{
\begin{array}{l}
-1,~-2,~\cdots ~~{\rm if}~~ c_{j}^2 x_{i}~>1,\\
{}\\
~0,~1,~2,~\cdots~~{\rm if}~~c_{j}^2 x_{i}~<1.
\end{array}\right.
\eqno(15)$$
\vskip .3cm
   In any case, up to a normalization constant  (14) can always be written as
\vskip .3cm
$$\vert N>=S^{+}_{0}S^{+}_{x_{1}}S^{+}_{x_{2}}\cdots S^{+}_{x_{N-1}}\vert 0>,\eqno(16)$$
\vskip .3cm
\noindent where 
\vskip .3cm
$$S^{+}_{x_{i}}=\sum_{j}{c_{j}\over{1-c^{2}_{j}x_{i}}}S^{+}_{j}.\eqno(17)$$
\vskip .3cm
\noindent While $x_{1},~x_{2},~\cdots, x_{N-1}$ are real or complex numbers satisfying the
relation
\vskip .3cm
$$-{1\over{2}}\sum^{p}_{j=1}\Omega_{j}c^{2}_{j}\alpha{1\over{1-\alpha y_{i}c^{2}_{j}}}
={1\over{y_{i}}}+
\sum_{k\neq i}{1\over{y_{i}-y_{k}}},~~i=1,~2,\cdots,~N-1,\eqno(18)$$
\vskip .3cm
\noindent with
$$\sum^{N-1}_{i=1}{1\over{y_{i}}}=1,\eqno(19)$$

\vskip .3cm
\noindent where
\vskip .3cm
$$y_{i}=x_{i}/\alpha,~~\alpha=-{2\over{h+2\Lambda_{1}}},~~h\equiv E/(-\vert G\vert).\eqno(20)$$
\vskip .3cm
\noindent Therefore, the coefficients $x_{i}$ (i=1,~2,$\cdots$, N-1) and eigenvalues of the
pairing energy $E\neq 0$ are simultaneously determined by the system of equations (18) and (19). 
\vskip .3cm
    While eigenvectors for $E=0$ can be expanded as follows 

$$\vert N>=\sum_{n_{i}}x_{1}^{n_{1}}x_{2}^{n_{2}}\cdots
x^{n_{N}}_{N}S^{+}_{n_{1}}S^{+}_{n_{2}}\cdots S^{+}_{n_{N}}\vert 0>,\eqno(21)$$

\noindent where  the restrictions on the integers $n_{i}$ are the same as those given by (15),
and the expansion coefficients $x_{i}$ are determined by the following set of equations

$$\sum_{j}\Omega_{j}{c^{2}_{j}\over{1-x_{1} c^{2}_{j}}}=0~~~{\rm for }~~N=1,\eqno(22)$$

$$\sum_{j}\Omega_{j}{c^{2}_{j}\over{1-x_{i} c^{2}_{j}}}=\sum_{k\neq i}{1\over{x_{i}-x_{k}}},~~~
i=1,~2,~\cdots,~N,~~{\rm for }~~N\geq 2.\eqno(23)$$

\vskip .3cm
 It is instructive to write down the first few energy eigenvalues and eigenstates.
\vskip .3cm
$$\hat{h}\vert 0>=0;\eqno(24)$$

$$\hat{h}\vert 1>=\sum_{j}\vert c_{j}\vert^{2}\Omega_{j}\vert 1>,~~\vert 1>=S^{+}_{0}\vert
0>,$$

$$\hat{h}\vert N=1,~\rho>~=~0,~~~~\vert N=1,~\rho>=\sum_{j}{c_{j}\over{1-x^{\rho}c^{2}_{j}}}S^{+}_{j}\vert
0>,~~\rho~=~1,~2,~\cdots,\eqno(25)$$

\noindent where $x_{\rho}$ is determined by (22);

$$\hat{h}\vert 2>= ~h\vert 2>,~~\vert 2>=S^{+}_{0}S^{+}_{x}\vert
0>,\eqno(26) $$
\vskip .3cm
\noindent where $h$ is one of the solutions of the equation
\vskip .3cm
$${1\over{2}}\sum^{p}_{j=1}\Omega_{j}c^{2}_{j}x{1\over{x c^{2}_{j}-1}}~=~1\eqno(27)$$
\vskip .3cm
\noindent where
\vskip .3cm
$$x=-{2\over{h+2\Lambda_{1}}},\eqno(28)$$
\vskip .3cm
$$\hat{h}\vert N=2,~\rho~>~=0,~~\rho~=~1,~2,~\cdots\eqno(29)$$

\noindent where

$$\vert
N=2,~\rho>=\sum_{j~j^{\prime}}{c_{j}c_{j^{\prime}}\over{(1-x^{\rho}_{1}c^{2}_{j})(1-x^{\rho}_{2}c^{2}_{j^{\prime}})}}
S^{+}_{j}S^{+}_{j^{\prime}}\vert 0>,\eqno(30)$$

\noindent and $x^{\rho}_{i}$ are determined by (23).
\vskip .3cm
   Energy levels of the generalized pairing interaction for
the $j~=~1/2,~3/2,$ and ~$5/2$ case with $c_{1/2}=\sqrt{0.1}$, $c_{3/2}=$ $\sqrt{0.3},$
and $c_{5/2}~=~\sqrt{0.6}$  are  shown in Fig. 1.
It can be seen from Table 1 that the energy level with the largest $c_j$ value for the
highest $j$ orbit is the lowest one for any fixed $N$. Finally, an example
of the lowest levels for given $N$ in the 5-th and 6-th shell, respectively, are shown in
Table 2. 

\newpage
\noindent {\bf Acknowledgment}
\vskip .5cm
   Supported  by the National Science Foundation through
Cooperative Agreement No. EPS-9550481 and Grant No. 96030006.
\vskip 1cm
\begin{centering}
{---------------------}\\
\end{centering}

\vskip 1cm
\begin{tabbing}
\=1111\=22222222222222222222222222222222222222222222222222222222222222222222
222222222222\=\kill\\
\>{[1]}\>{G. Racah, Phys. Rev. {\bf 62}, 438 (1942)}\\
\>{[2]}\>{J. Bardeen, L. N. Cooper, and J. R. Schrieffer, Phys. Rev. {\bf 108}, 1175 (1957)}\\
\>{[3]}\>{A. Bohr, B. R. Mottelson, and D. Pines, Phys. Rev. {\bf 110}, 936 (1958)}\\
\>{[4]}\>{S. T. Belyaev, Mat. Fys. Medd. {\bf 31}, 11 (1959)}\\
\>{[5]}\>{M. Randeria, J. M. Duan, and L. Y. Shieh, Phys. Rev. Lett. {\bf 62}, 981 (1989) }\\
\>{[6]}\>{S. Schmitt-Rink, C. M. Verma, and A. E. Ruckenstein, Phys. Rev. Lett. {\bf 63},
445 (1989)}\\
\>{[7]}\>{C. N. Yang, Phys. Rev. Lett. {\bf 63}, 2144 (1989)}\\
\>{[8]}\>{D. W. Cooper, J. S. Batchelder, and M. A. Taubenblatt, J. Coll. Int. Sci. {\bf 144},
201 (1991)}\\
\>{[9]}\>{M. Barranco, S. Hernandez, and R. J. Lombard, Z. Phys. {\bf D22}, 659(1992)}\\
\>{[10]}\>{J. Y. Zeng and T. S. Cheng, Nucl. Phys. {\bf A405}, 1 (1983); {\bf A411}, 49 (1984);
{\bf A414}, 253 (1984)}\\
\>{[11]}\>{A. K. Kerman, Ann. Phys. (N. Y.) {\bf 12}, 300 (1961)}\\
\>{[12]}\>{I. Talmi, Nucl. Phys. {\bf A172}, 1 (1971)}\\

\end{tabbing}
\newpage

\hoffset=-1.5cm

\noindent  {\bf Table 1.} The lowest energy levels for $j={\tiny~1/2,~3/2,~5/2}$ with:
\begin{tabbing}
{~~~~~~~~~~~~~~~~~}Case a: $c_{1/2}=\sqrt{0.1},~c_{3/2}=\sqrt{0.2},~c_{5/2}=\sqrt{0.7}$;\\
{~~~~~~~~~~~~~~~~~}Case b: $c_{1/2}=\sqrt{0.2},~c_{3/2}=\sqrt{0.1},~c_{5/2}=\sqrt{0.7}$;\\
{~~~~~~~~~~~~~~~~~}Case c: $c_{1/2}=\sqrt{0.2},~c_{3/2}=\sqrt{0.7},~c_{5/2}=\sqrt{0.1}$;\\
{~~~~~~~~~~~~~~~~~}Case d: $c_{1/2}=\sqrt{0.7},~c_{3/2}=\sqrt{0.2},~c_{5/2}=\sqrt{0.1}.$
\end{tabbing}
\small
\begin{tabbing}
\=111111111111111111111111111111111111111111111111111111111111111111111111111111111111111\=\kill\\
\>{------------------------------------------------------------------------------------------}\\
\=111111111\=22222222\=333333333333\=444444444444\=555555555555\=66666666666\=\kill\\
\>{}\>{}\>{~~~~~~~~~~~~~~~~~~$h=E/(-\vert G \vert$)}\\
\>{}\\
\>{~~~~~~~N}\>{}\>{a}\>{b}\>{c}\>{d}\\
\>{------------------------------------------------------------------------------------------}\\
\>{~~~~~~~0}\>{}\>{0}\>{0}\>{0}\>{0}\\
\>{~~~~~~~1}\>{}\>{2.6}\>{2.5}\>{1.9}\>{1.4}\\
\>{~~~~~~~2}\>{}\>{4.083}\>{3.866}\>{2.858}\>{2.136}\\
\>{~~~~~~~3}\>{}\>{4.735}\>{4.441}\>{3.278}\>{2.481}\\
\>{~~~~~~~4}\>{}\>{4.674}\>{4.377}\>{3.100}\>{2.320}\\
\>{~~~~~~~5}\>{}\>{4.084}\>{3.866}\>{2.858}\>{2.136}\\
\>{~~~~~~~6}\>{}\>{2.600}\>{2.500}\>{1.900}\>{1.400}\\
\>{------------------------------------------------------------------------------------------}\\
\end{tabbing}
\vskip .5cm
\noindent  {\bf Table 2.} Lowest energy levels for 5th and 6th shells. The parameters are 
chosen as follows:\\ 
$c_{1/2}=\sqrt{0.2},~c_{3/2}=\sqrt{0.3},~c_{5/2}=\sqrt{0.22},~c_{9/2}=\sqrt{0.28}$~ for 5th shell, 
~~~whereas $c_{1/2}=\sqrt{0.2},\\
c_{3/2}=\sqrt{0.14},~c_{5/2}=\sqrt{0.22},~c_{7/2}=\sqrt{0.28},~c_{11/2}=
\sqrt{0.16}$ for 
6th shell, respectively.\\
{--------------------------------------------------------------------------------------------------------------------------}\\
\begin{tabbing}
\=111111111111111111111111111111111\=2222222222222222222222\=\kill\\
\>{}\>{~~~~$(N,~h=E/(-\vert G\vert)$~)~~~~}\\
\end{tabbing}
{--------------------------------------------------------------------------------------------------------------------------}\\
\begin{tabbing}
\=1111111111\=222222222222\=333333333333\=444444444444\=555555555555
\=666666666666\=777777777777\=888888888888\=\kill\\
\>{5th shell}\>{(0, 0)}\>{(1, 2.860)}\>{(2, 5.192)}\>{(3, 6.999)}\>{(4, 8.283)}\>{(5, 9.046)}\\
\>{}\>{(6, 9.292)}\>{(7, 9.024)}\>{(8, 8.243)}\>{(9, 6.954)}\>{(10, 5.515)}\>{(11, 2.86)}\\
\>{}\\
\>{6th shell}\>{(0,0)}\>{(1, 3.220)}\>{(2, 6.014)}\>{(3, 8.387)}\>{(4, 10.345)}\>{(5, 11.893)}\\
\>{}\>{(6, 13.035)}\>{(7, 13.778)}\>{(8, 14.125)}\>{(9, 14.082)}\>{(10, 13.653)}\>{(11, 12.843)}\\
\>{}\>{(12, 11.655)}\>{(13, 10.096)}\>{(14, 8.167)}\>{(15, 5.874)}\>{(16, 3.220)}\\
\end{tabbing}
{--------------------------------------------------------------------------------------------------------------------------}\\

\newpage
\def\emline#1#2#3#4#5#6{%
\put(#1,#2){\special{em:moveto}}%
\put(#4,#5){\special{em:lineto}}}
\textwidth=16cm\textheight=24.5cm
\advance\textheight by -\headsep\advance\textheight by -\headheight
\oddsidemargin=0pt\evensidemargin=0pt\topmargin=0pt
\parindent=0pt\parskip=6pt plus 3pt minus 2pt
\raggedbottom\sloppy\pagestyle{empty}
\hoffset=.8cm
\unitlength 1.00mm
\linethickness{0.4pt}
\begin{centering}
\begin{picture}(135.33,155.11)
\put(10.33,150.00){\line(1,0){14.67}}
\put(7.00,150.00){\makebox(0,0)[cc]{}}
\put(40.00,150.00){\line(1,0){16.00}}
\put(36.00,150.00){\makebox(0,0)[cc]{2}}
\put(80.00,150.00){\line(1,0){15.00}}
\put(77.33,150.00){\makebox(0,0)[cc]{2}}
\put(110.00,150.00){\line(1,0){15.33}}
\put(40.67,125.00){\line(1,0){15.00}}
\put(36.00,125.00){\makebox(0,0)[cc]{}}
\put(80.00,132.67){\line(1,0){14.67}}
\put(80.33,127.33){\line(1,0){14.34}}
\put(80.00,109.00){\line(1,0){14.33}}
\put(110.00,140.33){\line(1,0){15.00}}
\put(110.33,135.00){\line(1,0){14.67}}

\put(110.00,126.00){\line(1,0){15.00}}
\put(110.00,116.00){\line(1,0){15.00}}
\put(110.00,102.33){\line(1,0){15.00}}
\put(2.33,90.00){\framebox(133.00,65.11)[cc]{}}
\put(2.33,150.00){\line(1,0){2.34}}
\put(2.33,140.00){\line(1,0){2.00}}
\put(2.33,130.00){\line(1,0){2.00}}
\put(2.33,120.00){\line(1,0){1.67}}
\put(2.33,120.00){\line(1,0){2.34}}
\put(2.33,120.00){\line(1,0){2.00}}
\put(2.33,110.00){\line(1,0){2.00}}
\put(2.33,100.00){\line(1,0){2.00}}
\put(-2.00,150.00){\makebox(0,0)[cc]{0}}
\put(-2.00,140.00){\makebox(0,0)[cc]{-1}}
\put(-2.00,130.00){\makebox(0,0)[cc]{-2}}
\put(-2.00,120.00){\makebox(0,0)[cc]{-3}}
\put(-2.00,110.00){\makebox(0,0)[cc]{-4}}
\put(-2.00,100.00){\makebox(0,0)[cc]{-5}}
\put(-14.00,125.00){\makebox(0,0)[cc]{$E/\vert G\vert$}}
\put(18,95){\makebox(0,0){$N=0$}}
\put(48,95){\makebox(0,0){$N=1$}}
\put(86,95){\makebox(0,0){$N=2$}}
\put(118,95){\makebox(0,0){$N=3$}}
\end{picture}
\vskip -8cm
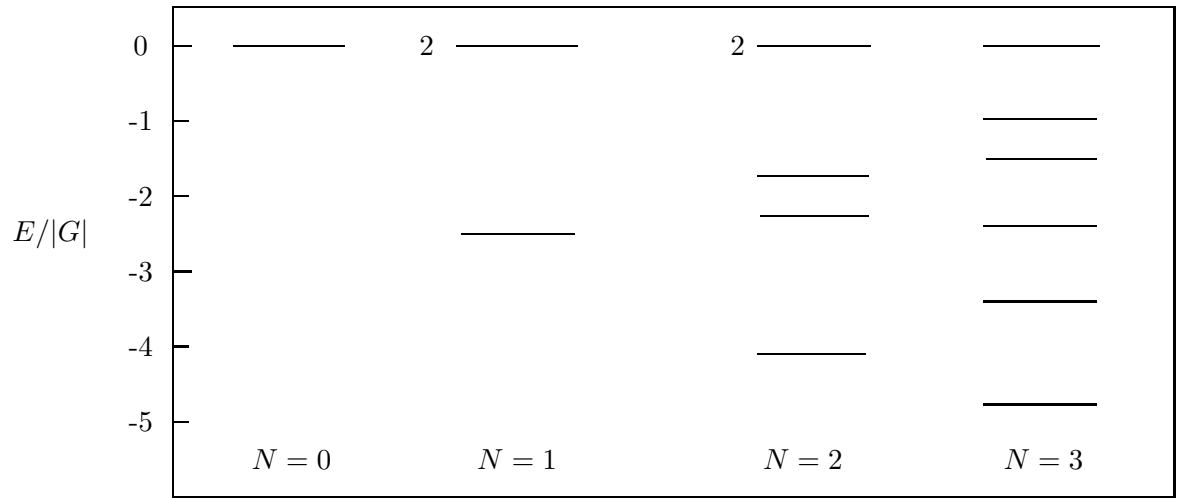
{\bf Fig. 1.} An example of excited energy levels for the generalized pairing interaction with
$N\leq 3$, where the number on the left indicates the degeneracy when
it is greater than one. 
\end{centering}
\end{document}